\documentclass{article}
\usepackage{graphicx}
\usepackage[margin=1in]{geometry}
\usepackage{amsmath}
\usepackage{amssymb}
\usepackage{amsthm}
\usepackage{accents}
\usepackage[maxfloats=128]{morefloats}
\usepackage[breaklinks,hidelinks]{hyperref}

\newtheorem{theorem}{Theorem}[]

\newtheorem{remark1}[theorem]{Remark}
\newenvironment{remark}{\begin{remark1} \rm}{\end{remark1}}

\title{Simulating single-coil MRI from the responses of multiple
       coils\thanks{``MRI'' abbreviates ``magnetic resonance imaging''}}
\author{Mark Tygert and Jure Zbontar}

\begin{document}

\maketitle

\begin{abstract}
We convert the information-rich measurements of parallel and phased-array MRI
into noisier data that a corresponding single-coil scanner could have taken.
Specifically, we replace the responses from multiple receivers
with a linear combination that emulates the response from only a single,
aggregate receiver, replete with the low signal-to-noise ratio
and phase problems of any single one of the original receivers
(combining several receivers is necessary, however,
since the original receivers usually have limited spatial sensitivity).
This enables experimentation in the simpler context of a single-coil scanner
prior to development of algorithms for the full complexity
of multiple receiver coils.
\end{abstract}

\section{Introduction}

Typical modern MRI scanners (such as all those represented
in the fastMRI data of~\cite{fastMRI}) include {\it multiple} receivers
--- so-called ``coils'' of conductors --- for making measurements.
In this brief yet self-contained technical note,
we seek to conveniently simulate the response of a single-coil machine,
complete with errors that are reasonably representative
of the measurement errors which a real scanner would make,
given only the responses from a multi-coil, ``parallel-imaging'' machine.
The fastMRI competition issues two challenges --- single-coil and multi-coil
--- under the expectation that the single-coil track could provide
a simplified stepping-stone toward the full multi-coil challenge.

The aim of fastMRI is compressed sensing ---
the accelerated acquisition of image reconstructions
by taking fewer measurements than classical signal processing would need
to attain the same resolution; compressed sensing yields full-resolution
or ``superresolved'' reconstructions based on so-called ``undersampled''
measurements, measurements acquired at lower than the Nyquist rate
required to reconstruct any arbitrary signal for a given bandwidth.
Two approaches to compressed sensing are especially popular:
(1) optimization for individual images and
(2) machine learning over collections of images.
Combinations of the two approaches are also possible.
Techniques representative of the first approach (via optimization)
include Fourier reconstruction with total-variation regularizers
(or related regularizers, such as the sum of the absolute values
of wavelet coefficients); representative of the second approach
(via machine learning) is statistical modeling with deep neural networks.
The terminology, ``compressed sensing,'' ``compressive sensing,''
and ``compressive sampling,'' is due to~\cite{candes2005decoding}
and~\cite{donoho2006compressed}; they took the first approach
(via optimization). The second approach (machine learning)
is the focus of~\cite{fastMRI}.

Complicating the development of compressed sensing for MRI
is the longstanding focus on leveraging multiple receiver coils
in the design of MRI scanners, as reviewed by~\cite{ohliger-sodickson}.
All scanners of~\cite{fastMRI} are multi-coil by default,
designed and optimized for multi-coil measurements.
However, compressed sensing is more complex with multi-coil data.
Distilling the multi-coil challenge into a single-coil stepping-stone
could lower barriers to entry into research on compressed sensing for MRI
and accelerate the development of accelerated MRI acquisitions,
by teasing apart the fundamental mathematical issues
from the multi-coil complications.

Thus, we need to simulate a single-coil scanner with sufficient fidelity
to actual errors in measurement, using only multi-coil data.
As hoped (and illustrated with numerical examples
in Section~\ref{numex} below), our single-coil emulation does indeed yield
phase similar to the notorious phase problems
tackled by~\cite{lee-pauly-nishimura} and others.
Encouragingly, the method
of~\cite{setsompop-wald-alagappan-gagoski-adalsteinsson} is similar to ours
and yielded highly successful results in a related setting
(though, admittedly, our mathematical motivation, practical application,
optimization procedure and targets, and detailed objective differ).
Section~\ref{esc} now introduces our simulation of a single-coil scanner.

\section{Emulated single-coil}
\label{esc}

We propose using a linear combination of the responses from multiple coils
for the emulated single-coil (ESC) response.
We least-squares fit the complex-valued coefficients in the linear combination
to the ``ground-truth'' reconstruction,
estimating the ground truth using the canonical full multi-coil reconstruction,
the root-sum-square (RSS) of~\cite{roemer-edelstein-hayes-souza-mueller}.
Most notably, linearly combining the raw coil responses sums
complex-valued fluxes directly, rather than summing nonnegative energies
as in the RSS.
This simulates a single-coil scanner for the reasons discussed next.

\subsection{Motivation}

\begin{figure}
\begin{center}
\parbox{.45\textwidth}{\includegraphics[width=.45\textwidth]{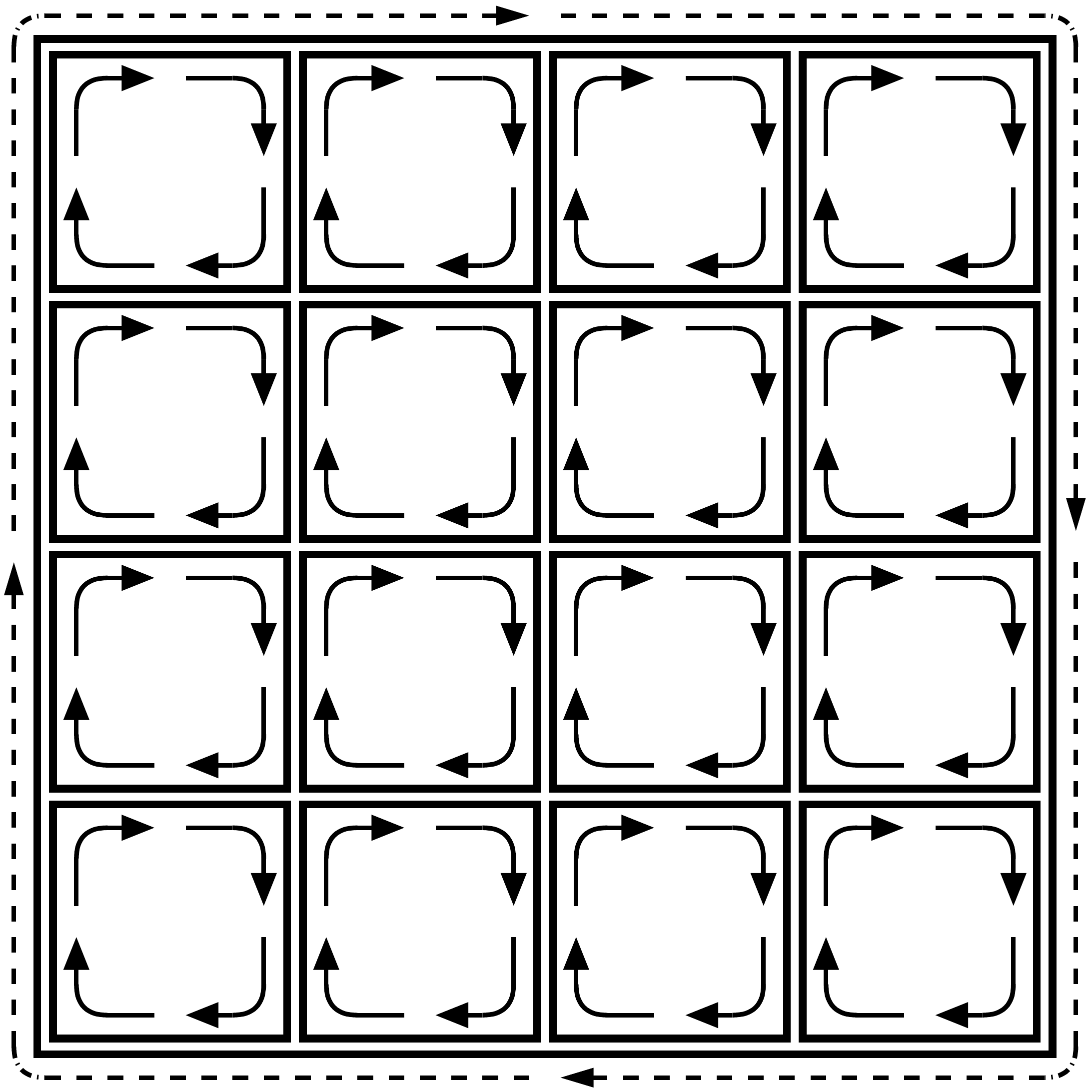}}
\end{center}
\caption{The wide solid lines represent coils of conductors
--- 16 small ones and 1 large one. The arrows indicate circulation of current
through the conductors (with solid lines for the small coils and dashed lines
for the large coil). The currents in the small coils would sum to induce
the indicated current in the large coil (flux through the small coils would be
equivalent to flux through the large coil), neglecting mutual inductances ---
and actual designs of coil arrays tend to minimize mutual inductances
(as per~\cite{vaughan-griffiths}).}
\label{Stokes}
\vspace{.25in}
\end{figure}

As seen in Figure~\ref{Stokes}, the flux through a single coil is physically
the sum of the flux through multiple coils forming a disjoint partition
of the single coil that loops around the multiple coils (provided that
the multiple coils are all in phase and have the same gain, while neglecting
the mutual inductance that coil design typically tries to minimize,
as per~\cite{vaughan-griffiths}).
This is much like in the proof of Stokes' Theorem
(which also uses Figure~\ref{Stokes}).
Since we do not know the correct phase offset and gain calibrations,
which amount to multiplying a whole coil's measurements
by the same complex number, we can fit the complex number
so as to match the ground-truth RSS reconstruction as well as possible.

As described, for example, by~\cite{vaughan-griffiths}, many MRI machines
use ``bird-cage'' or other volumetric coils rather than surface coils
in a flat plane.
A phased ``single-coil'' array is often a cylindrical array of coils
with capacitative (or inductive) coupling between the coils.
This coupling effectively introduces phase offsets
between the multiple conductors forming the ``single-coil'' bird-cage.
Our linear combination accounts for phase offsets and is physically realizable
and similar to bird cages used in practice.
The use of coils selecting for circular polarization
(including the quadrature pairs, butterflies, figure eights, etc.\
described by~\cite{vaughan-griffiths}) is an additional complication,
implicitly resolved by how the least-squares fit of the linear combination
to the ground truth decides to combine the coils.
The following subsection formalizes the mathematical details.

\subsection{Mathematical formulation}

Given data $A$ from multiple coils,
we define $k$ to be the number of coils,
$l$ to be the number of slices in a volume of cross-sectional images,
$m$ to be the height of each slice,
and $n$ to be the width of each slice;
$A$ then consists of an $(nml) \times k$ matrix with complex-valued entries.
More precisely, the images in $A$ are cropped inverse Fourier transforms
of the original two-dimensional Fourier domain (``$k$-space'') measurements,
cropping to the center $m \times n$ block of pixels.
We define $b$ to be the $(nml) \times 1$ column vector
containing ``ground-truth'' reconstructions from the full multi-coil data,
such as the RSS of~\cite{roemer-edelstein-hayes-souza-mueller};
specifically, each entry of $b$ is the Euclidean norm
of the corresponding $1 \times k$ row in $A$.
We then calculate the $k \times 1$ column vector $x$
with complex-valued entries minimizing
\begin{equation}
\label{obfun}
\bigl\| \sqrt{|Ax|} - \sqrt{|b|} \bigr\|_2^2,
\end{equation}
where $\|\cdot\|_2$ is the Euclidean norm,
$\sqrt{\cdot}$ takes the square root entrywise,
and $|\cdot|$ takes the absolute value entrywise,
so that $|b|$ is in fact equal to $b$,
as the entries of $b$ are nonnegative.
The distance $\|\sqrt{|Ax|}-\sqrt{|b|}\|_2$ is known as the Hellinger metric
of~\cite{hellinger}; the square roots amplify entries of small magnitude
and attenuate entries of large magnitude. Furthermore,
$\sqrt{|Ax|} - \sqrt{|b|} = (|Ax| - |b|)/(\sqrt{|Ax|} + \sqrt{|b|})$,
so the Hellinger distance is more like a measure of relative errors
than the conventional Euclidean distance  $\bigl\| |Ax| - |b| \bigr\|_2$.
Our estimate of $|b|$ is $|Ax|$, where $x$ minimizes~(\ref{obfun}).
Minimizing~(\ref{obfun}) is far from the only possibility;
minimizing~(\ref{obfun}) performed well in our numerical experiments
and facilitates the nonlinear optimization required.

To solve the nonlinear least-squares problem minimizing~(\ref{obfun}),
we tried three different methods. The three methods (all reviewed,
for example, by~\cite{nocedal-wright}) are $\{1\}$ gradient descent,
$\{2\}$ the Levenberg-Marquardt variant of Gauss-Newton iterations,
and $\{3\}$ the LBFGS (limited-memory Broyden-Fletcher-Goldfarb-Shanno) version
of quasi-Newton iterations.
For all, we started the iterations with the solution
to the linear least-squares problem minimizing
\begin{equation}
\label{lls}
\| Ax - b \|_2^2;
\end{equation}
the objective function in~(\ref{obfun}) is the same as in~(\ref{lls}) aside
from taking the square roots of the absolute values of the entries
of $Ax$ and of $b$.

Gradient descent was the slowest, and its best step size varied dramatically.
Levenberg-Marquardt and LBFGS both worked well,
but LBFGS was an order of magnitude faster on the wall clock.
Levenberg-Marquardt requires forming a large system matrix explicitly,
whereas LBFGS uses only matrix-vector multiplications
and so can be ``matrix-free,'' not requiring the explicit formation
of the system matrix.
The results reported below come from LBFGS.
We used the implementations in SciPy of~\cite{scipy} for all experiments.

\begin{remark}
Anuroop Sriram points out that the mean of the noise on the ``ground-truth''
RSS will be strictly positive in regions of the image where the actual object
being imaged is absent, as the RSS is always nonnegative.
Thus, when we fit the linear combination of the coil responses,
we should try to match the RSS only where the RSS is substantially above
a threshold that is significantly above the noise level. For the current data
from fastMRI of~\cite{fastMRI}, essentially the entire cropped image
of interest meets this criterion, so we try to match the full cropped image.
\end{remark}

\section{Numerical examples}
\label{numex}

We compare the emulated single-coil (ESC) method of Section~\ref{esc}
with ESPIRiT of~\cite{espirit}
and the leading singular vector of the singular value decomposition (SVD)
of $A$ from~(\ref{obfun}),
as in the ``eigencoil'' or ``coil compression'' reviewed
by~\cite{ohliger-sodickson}.
ESC could be regarded as a special kind of the ``virtual-coil'' methods
discussed by~\cite{ohliger-sodickson}.
We also display the root-sum-square (RSS)
of~\cite{roemer-edelstein-hayes-souza-mueller}
for reference as our best estimate of the ground truth.

\begin{figure}
\centerline{\parbox{.49\textwidth}{\includegraphics[width=.49\textwidth]{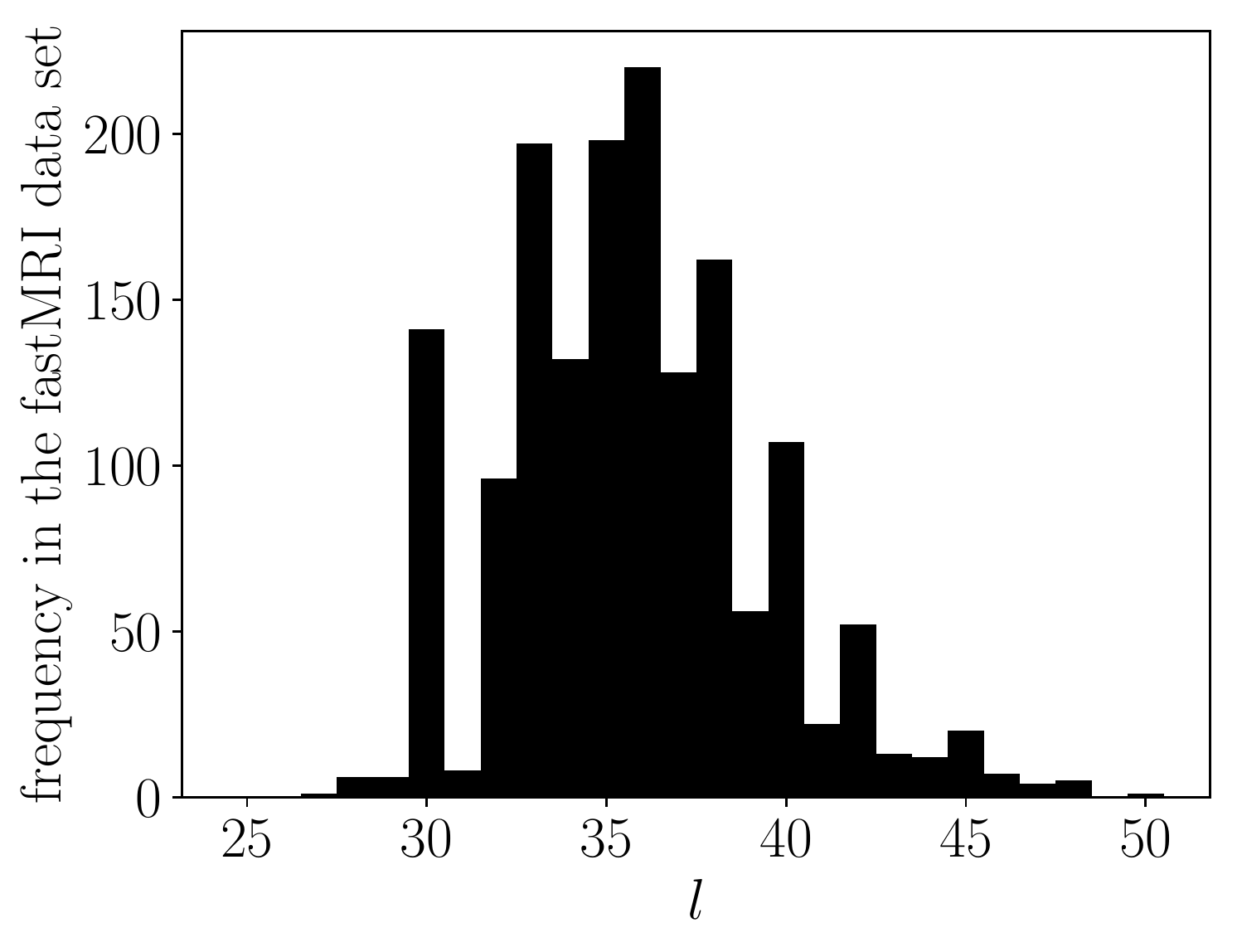}}}
\caption{number of examples in the fastMRI data set
         of~\cite{fastMRI} for each possible value of $l$
         (the parameter $l$ is the number of cross-sectional slices)}
\label{slices}
\vspace{.5in}
\end{figure}

For our data --- scans of 1594 knees from~\cite{fastMRI} ---
the parameters from Section~\ref{esc} take the values $k = 15$ receiver coils,
$m = 320$ rows and $n = 320$ columns in a cross-sectional image,
with Figure~\ref{slices} specifying $l$, the number of cross-sectional slices.
We used a memory of 10 vectors in the limited-memory BFGS
quasi-Newton iterations to minimize~(\ref{obfun})
starting from the linear least-squares solution minimizing~(\ref{lls}).

\begin{figure}
\parbox{\textwidth}{\includegraphics[width=\textwidth]{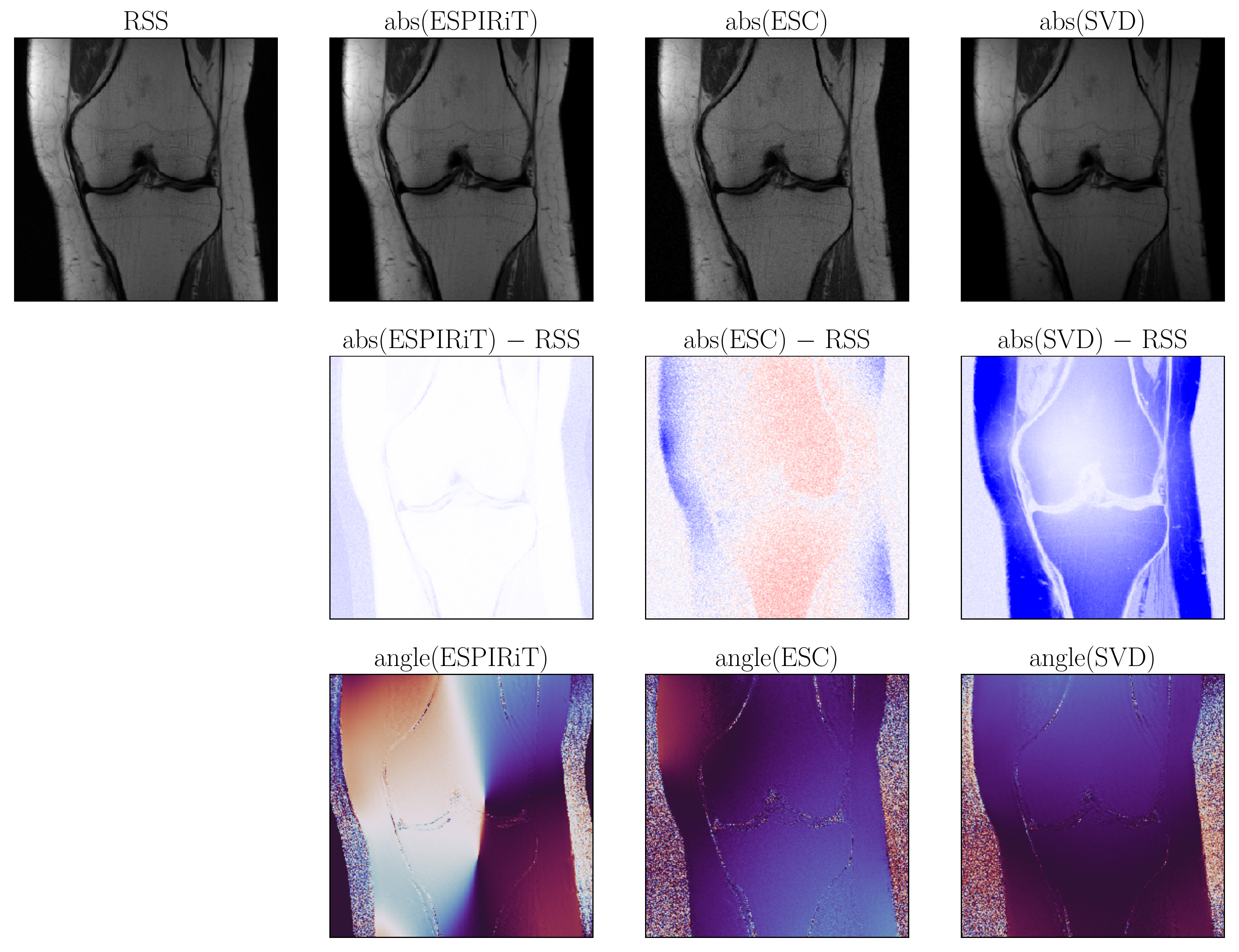}}
\caption{first randomly chosen example; ``abs'' refers to the absolute value
of a complex number and ``angle'' refers to the phase}
\label{jzb1}
\end{figure}

\begin{figure}
\parbox{\textwidth}{\includegraphics[width=\textwidth]{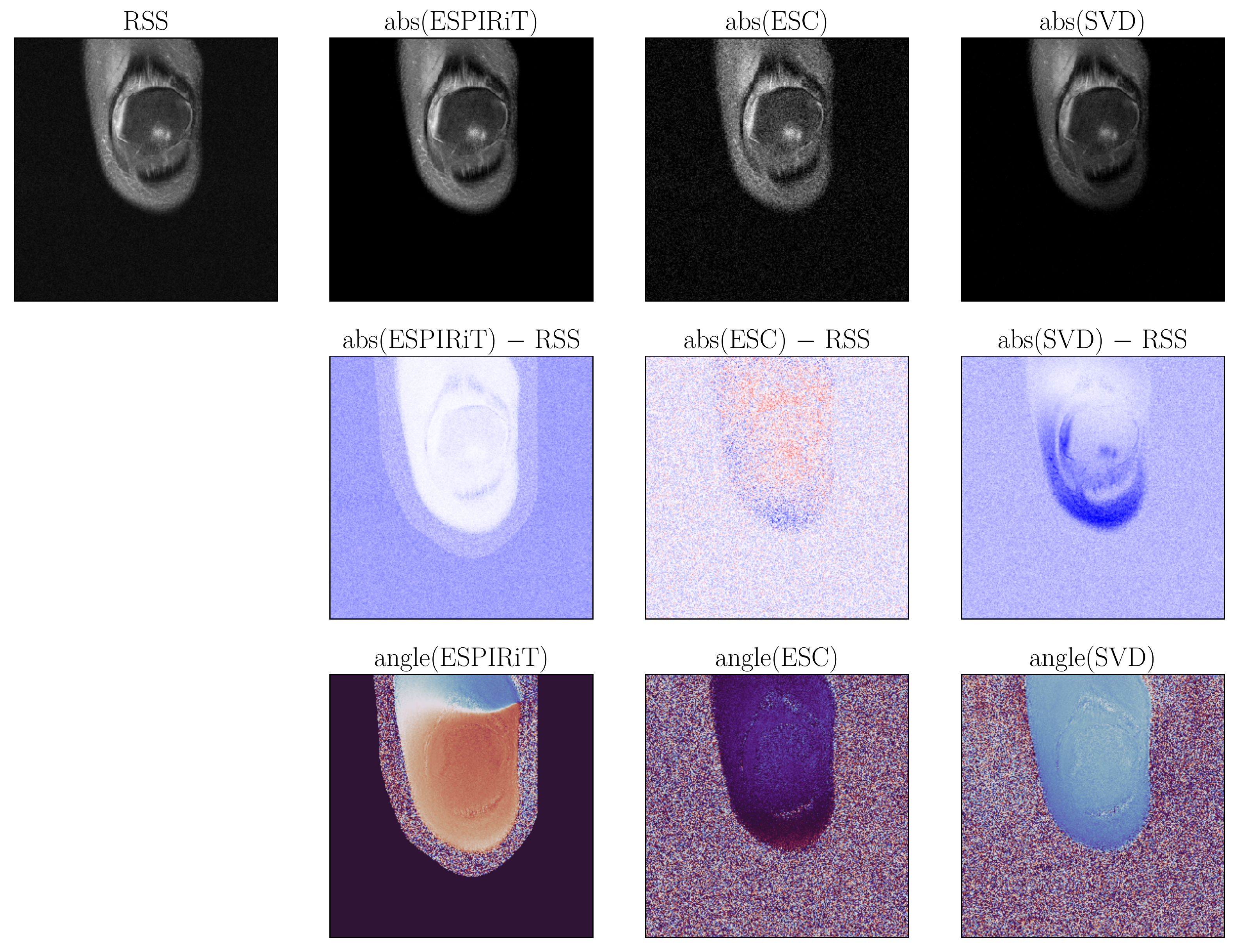}}
\caption{second randomly chosen example; ``abs'' refers to the absolute value
of a complex number and ``angle'' refers to the phase}
\label{jzb2}
\end{figure}

\begin{figure}
\parbox{\textwidth}{\includegraphics[width=\textwidth]{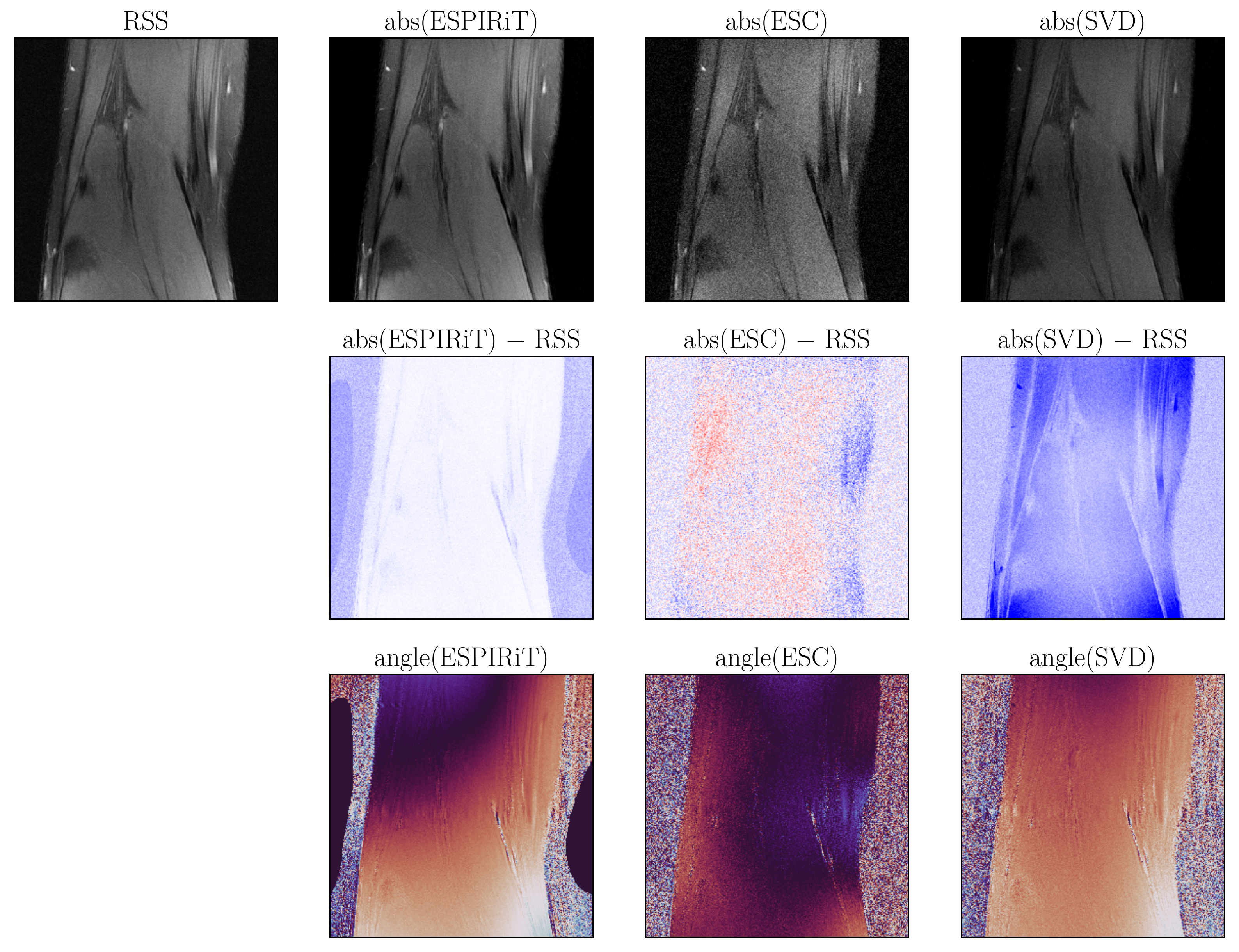}}
\caption{third randomly chosen example; ``abs'' refers to the absolute value
of a complex number and ``angle'' refers to the phase}
\label{jzb3}
\end{figure}

\begin{figure}
\parbox{\textwidth}{\includegraphics[width=\textwidth]{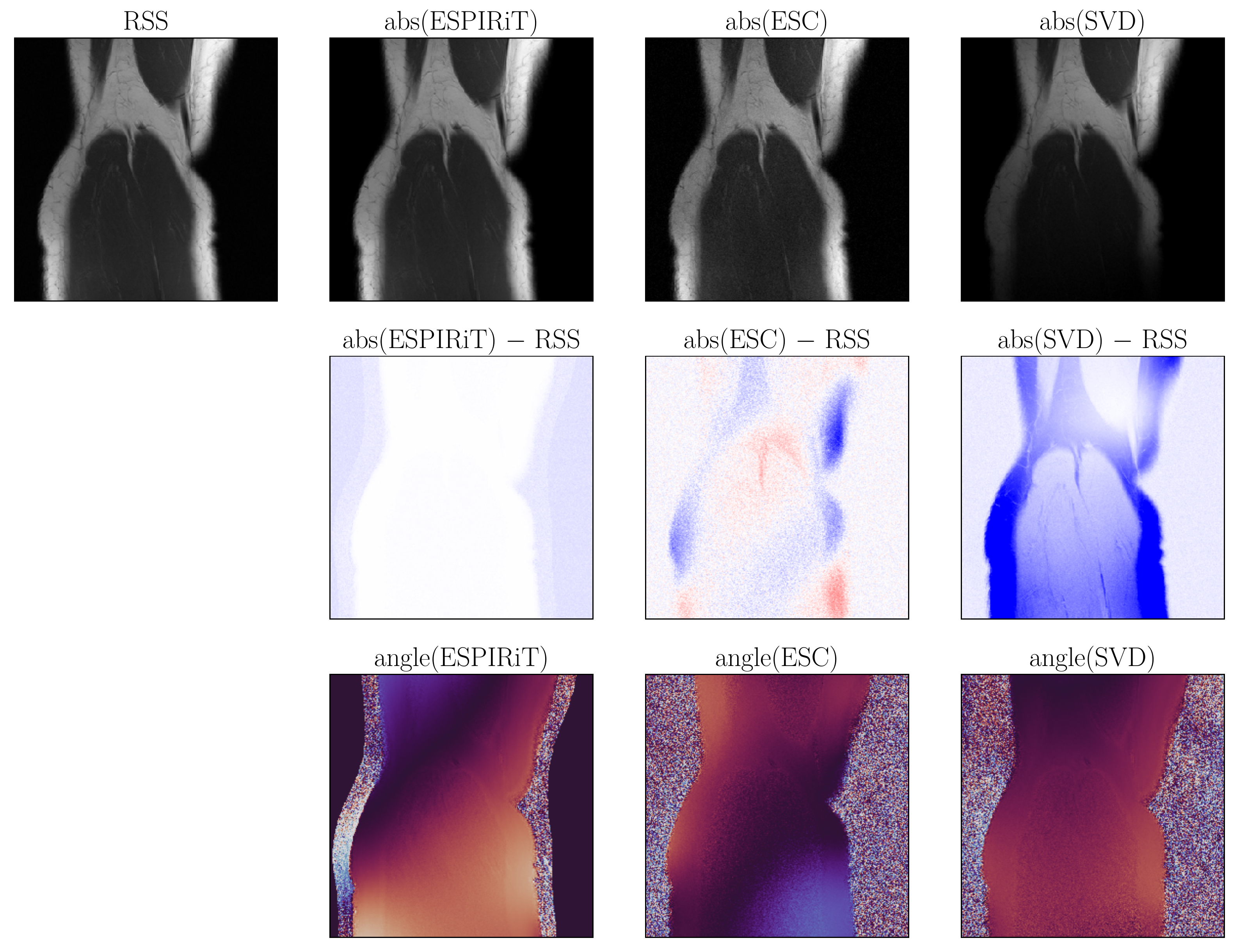}}
\caption{fourth randomly chosen example; ``abs'' refers to the absolute value
of a complex number and ``angle'' refers to the phase}
\label{jzb4}
\end{figure}

\begin{figure}
\parbox{\textwidth}{\includegraphics[width=\textwidth]{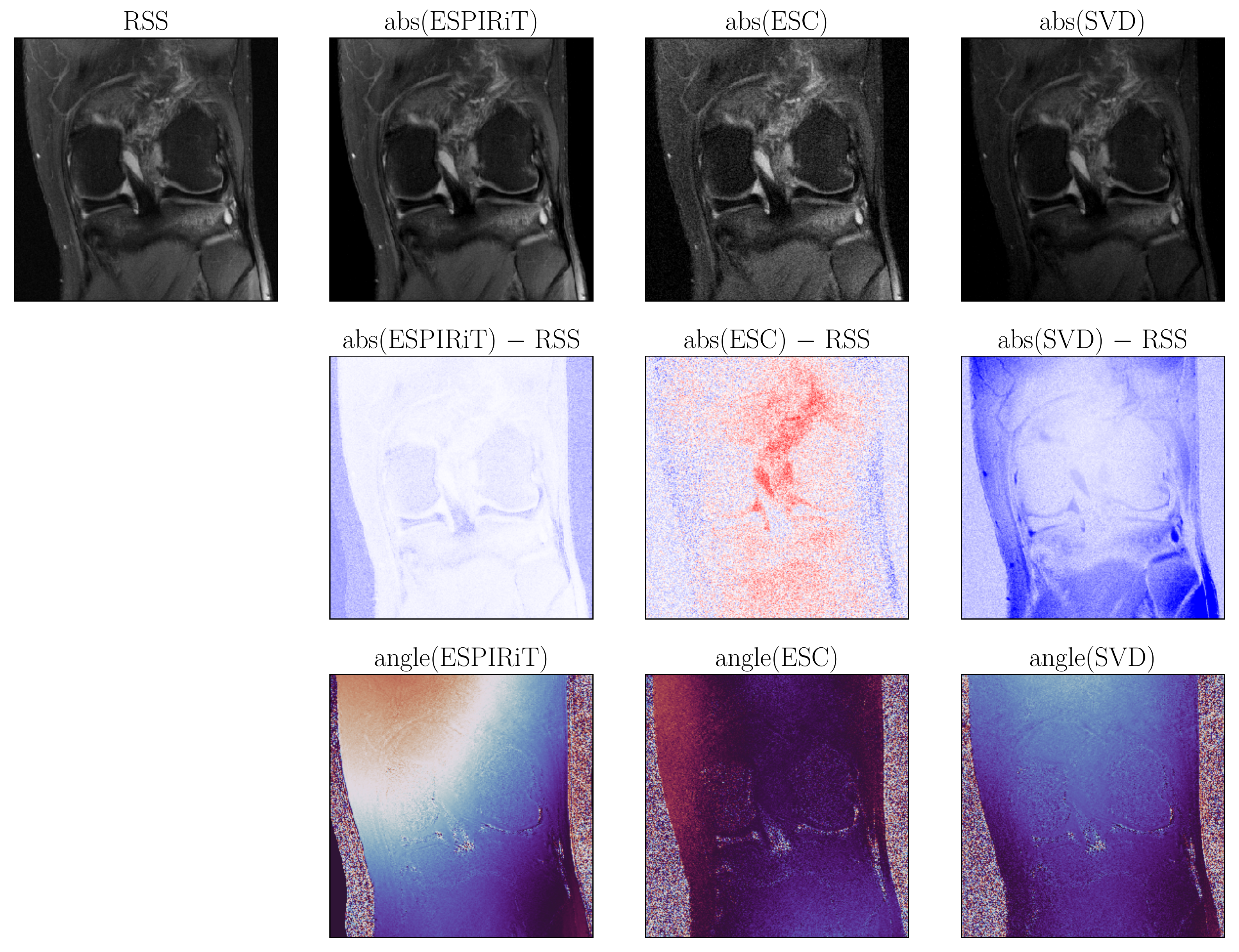}}
\caption{fifth randomly chosen example; ``abs'' refers to the absolute value
of a complex number and ``angle'' refers to the phase}
\label{jzb5}
\end{figure}

\begin{figure}
\parbox{\textwidth}{\includegraphics[width=\textwidth]{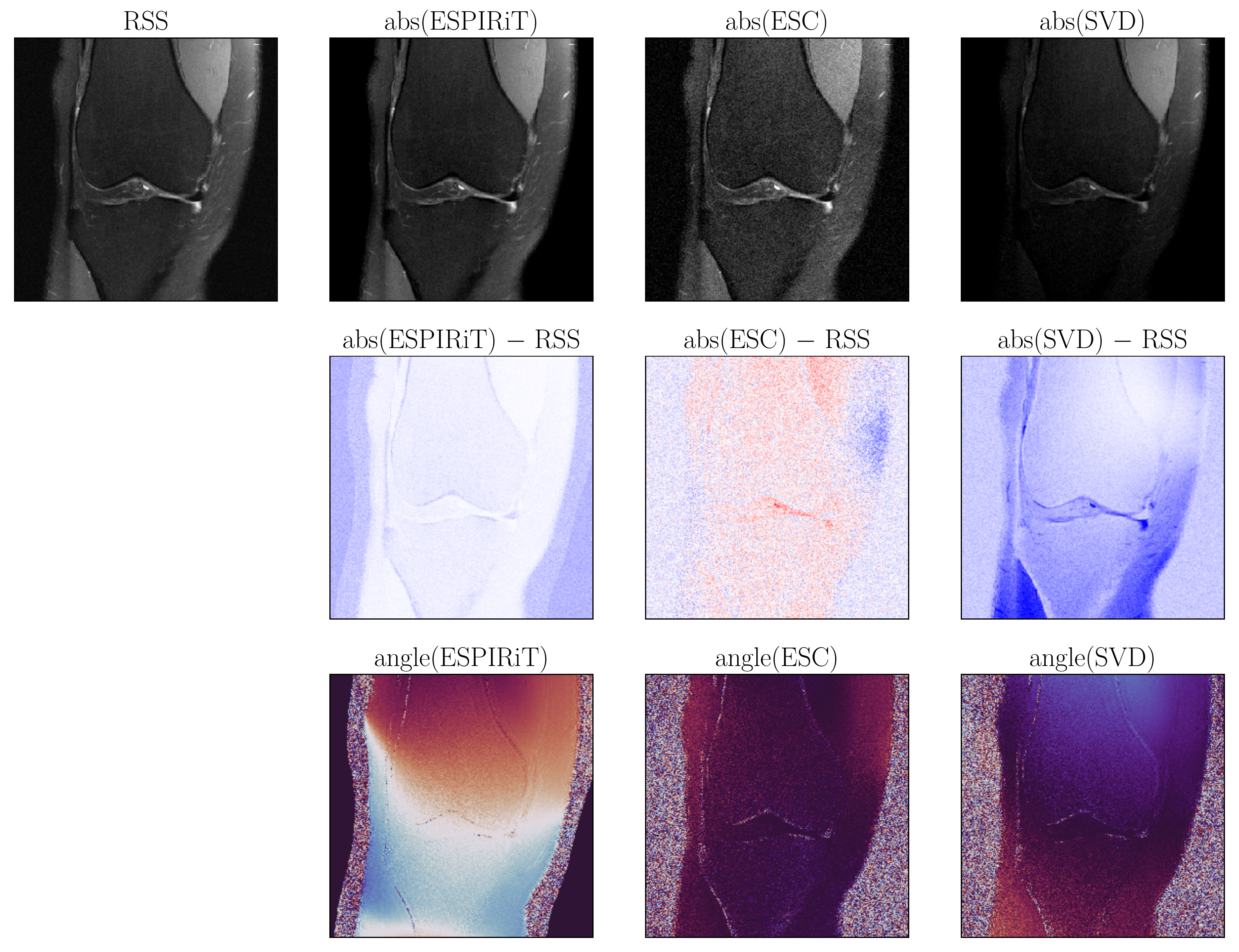}}
\caption{sixth randomly chosen example; ``abs'' refers to the absolute value
of a complex number and ``angle'' refers to the phase}
\label{jzb6}
\end{figure}

\begin{figure}
\begin{center}
\parbox{.30\textwidth}{\includegraphics[width=.30\textwidth]{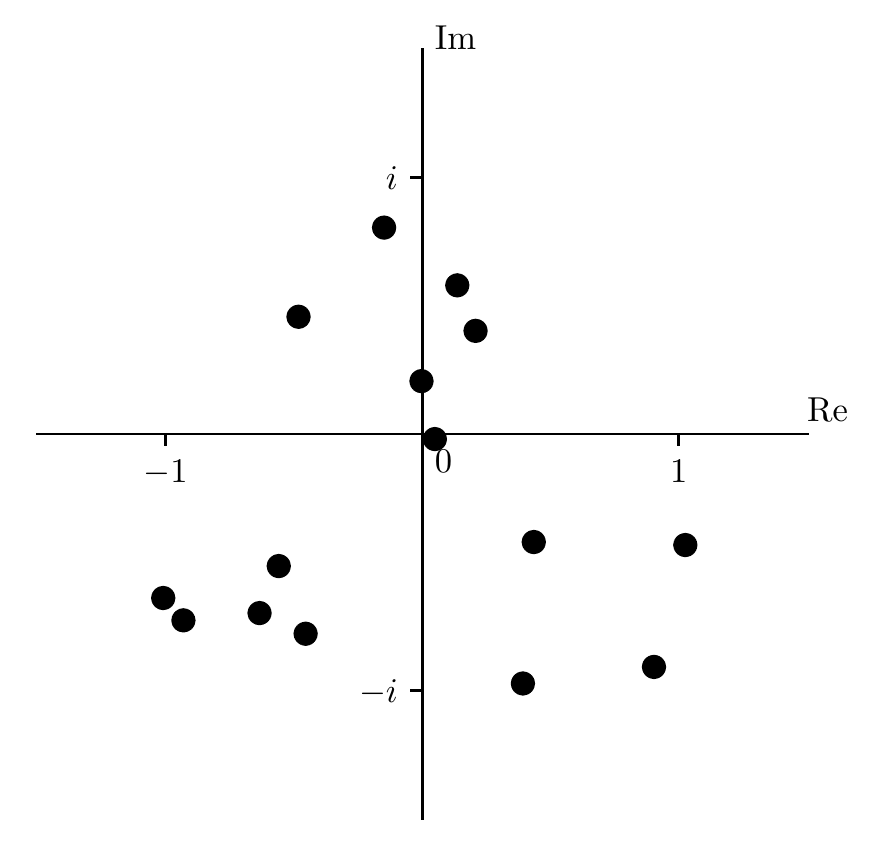}}
\end{center}
\caption{entries of $x$ minimizing~(\ref{obfun}) corresponding
to the data in Figure~\ref{jzb1}}
\label{plt1}
\end{figure}

\begin{figure}
\begin{center}
\parbox{.30\textwidth}{\includegraphics[width=.30\textwidth]{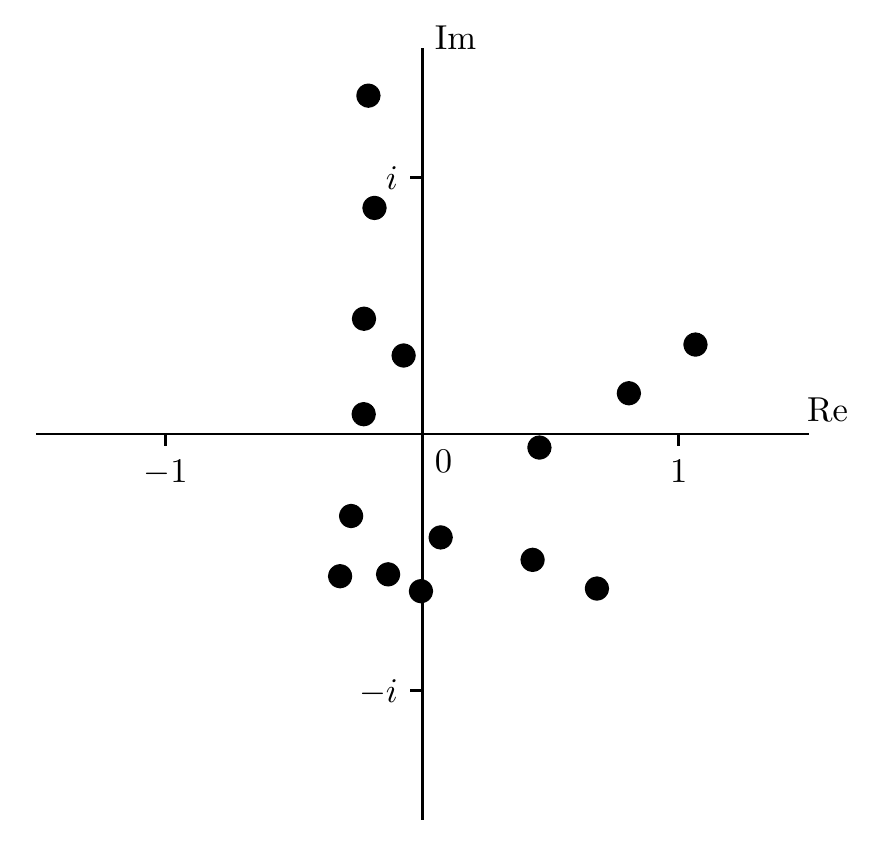}}
\end{center}
\caption{entries of $x$ minimizing~(\ref{obfun}) corresponding
to the data in Figure~\ref{jzb2}}
\label{plt2}
\end{figure}

\begin{figure}
\begin{center}
\parbox{.30\textwidth}{\includegraphics[width=.30\textwidth]{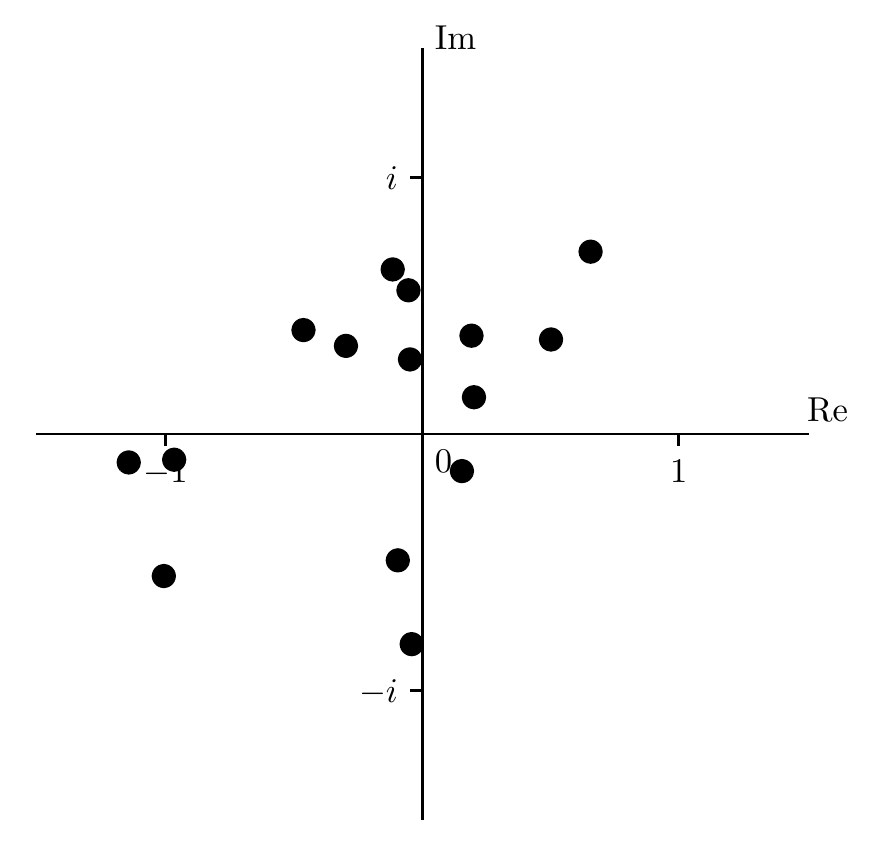}}
\end{center}
\caption{entries of $x$ minimizing~(\ref{obfun}) corresponding
to the data in Figure~\ref{jzb3}}
\label{plt3}
\end{figure}

\begin{figure}
\begin{center}
\parbox{.30\textwidth}{\includegraphics[width=.30\textwidth]{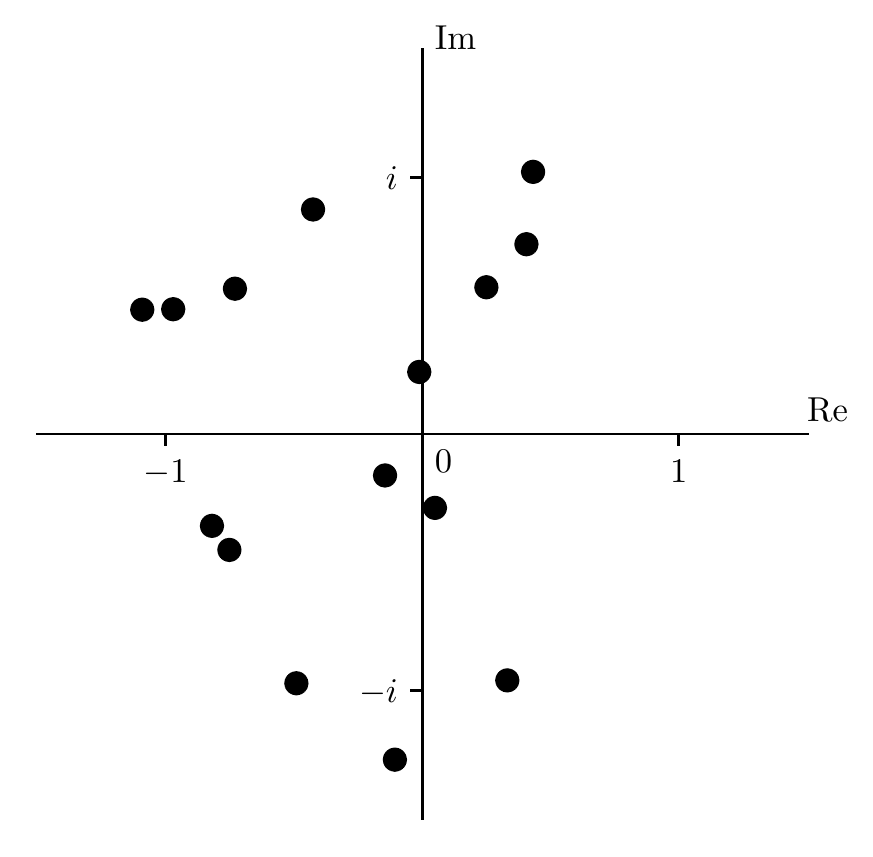}}
\end{center}
\caption{entries of $x$ minimizing~(\ref{obfun}) corresponding
to the data in Figure~\ref{jzb4}}
\label{plt4}
\end{figure}

\begin{figure}
\begin{center}
\parbox{.30\textwidth}{\includegraphics[width=.30\textwidth]{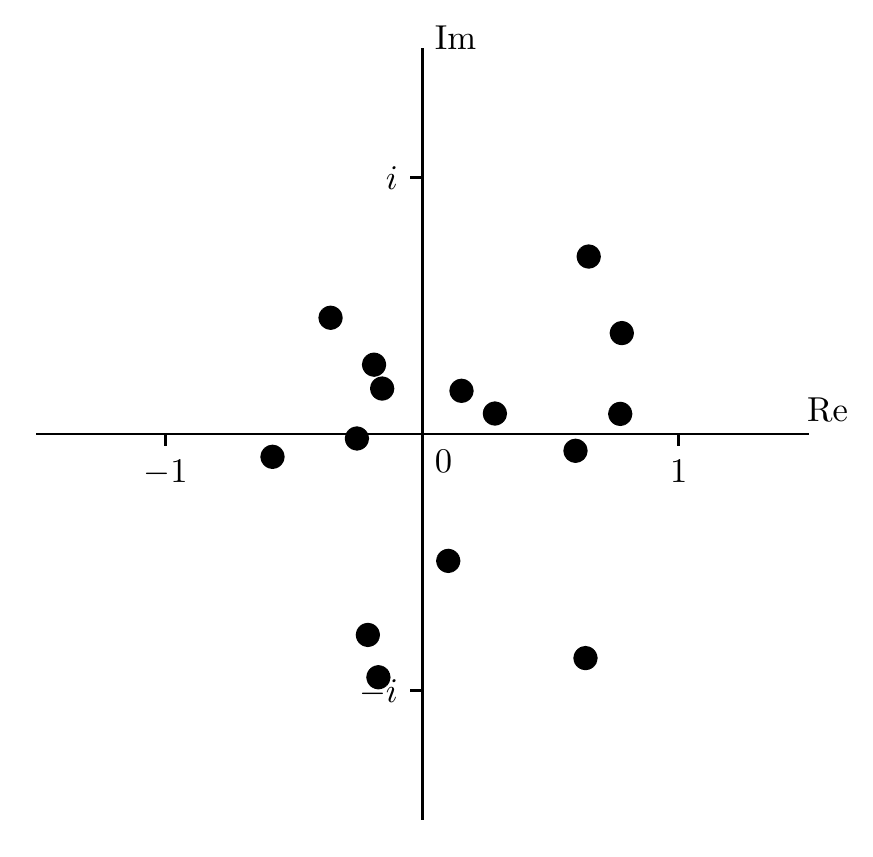}}
\end{center}
\caption{entries of $x$ minimizing~(\ref{obfun}) corresponding
to the data in Figure~\ref{jzb5}}
\label{plt5}
\end{figure}

\begin{figure}
\begin{center}
\parbox{.30\textwidth}{\includegraphics[width=.30\textwidth]{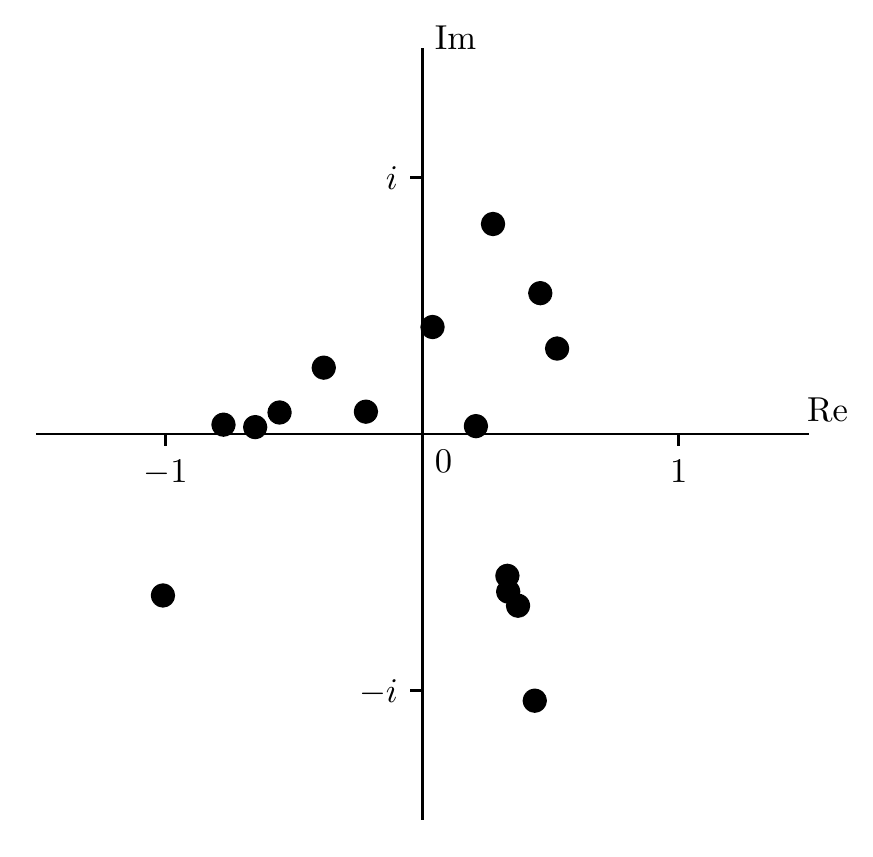}}
\end{center}
\caption{entries of $x$ minimizing~(\ref{obfun}) corresponding
to the data in Figure~\ref{jzb6}}
\label{plt6}
\end{figure}

In Figures~\ref{jzb1}--\ref{jzb6}, the first row shows (the magnitude of)
the reconstructions. The second row shows the difference between
the RSS ground truth and the magnitude of each method's reconstruction.
The third row shows the phase of the reconstructions
(estimates of the RSS ground truth discard the phase).
Figures~\ref{plt1}--\ref{plt6} display the entries of the solution $x$
minimizing~(\ref{obfun}), where the ESC reconstruction is $Ax$.

Figures~\ref{jzb1}--\ref{jzb6} show that the SVD-based compression
to a single eigenmode fails miserably in all respects
--- compressing to only one eigenmode is apparently too drastic.
The reconstruction from ESPIRiT matches the ground-truth RSS well,
and has absolutely no noise away from the actual object being imaged,
unlike real measurements from actual coils.
Of course, ESPIRiT makes no attempt to simulate a single-coil scanner.

Figures~\ref{jzb1}--\ref{jzb6} show that the reconstruction from ESC
matches the ground-truth RSS reasonably well, too, but is noisy.
The noise on the reconstruction from ESC
is actually desirable for simulating a single-coil scanner
(since a single-coil scanner is likely to be noisier than the ground truth
estimated from the full multi-coil data).
This matches our expectation that ESC emulates a physically realizable process
for converting a multi-coil scanner into a single-coil scanner.
The scanner which ESC simulates is genuinely physically realizable,
even though it only approximates single-coil scanners
that have actually been manufactured.

\begin{figure}
\centerline{\parbox{.8\textwidth}{\includegraphics[width=.8\textwidth]{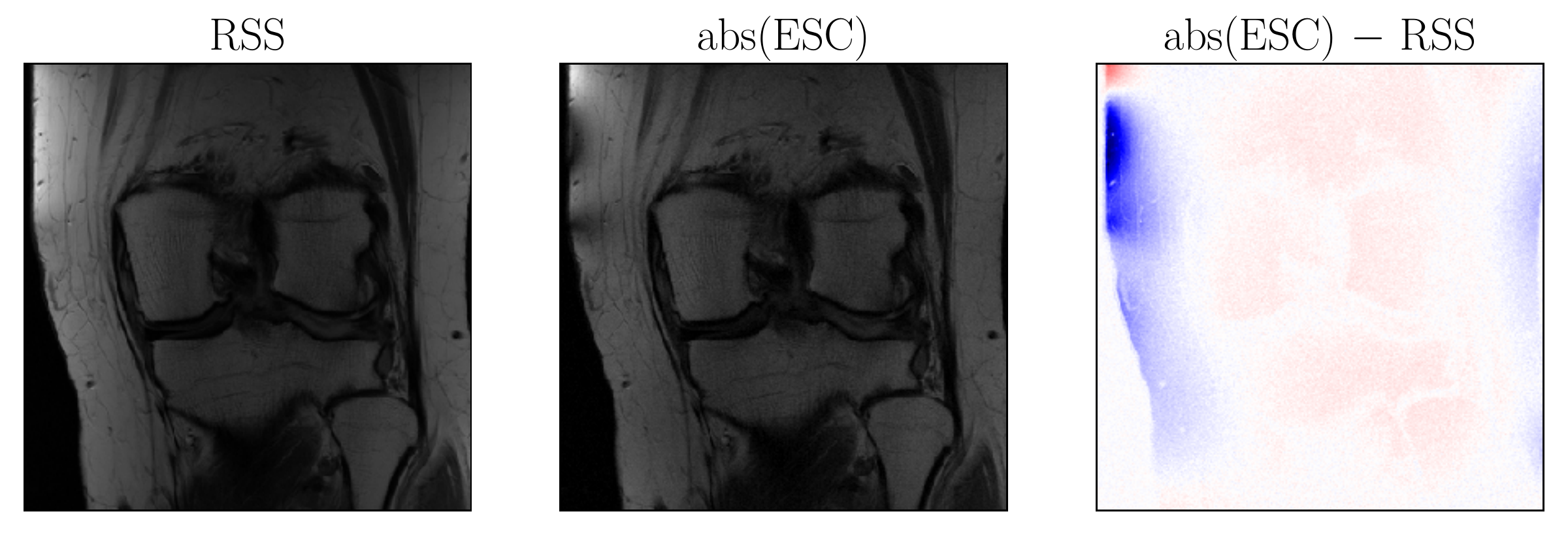}}}
\caption{``worst'' abs(ESC)$-$RSS;
``abs'' refers to the absolute value of a complex number}
\label{err1}
\end{figure}

\begin{figure}
\centerline{\parbox{.8\textwidth}{\includegraphics[width=.8\textwidth]{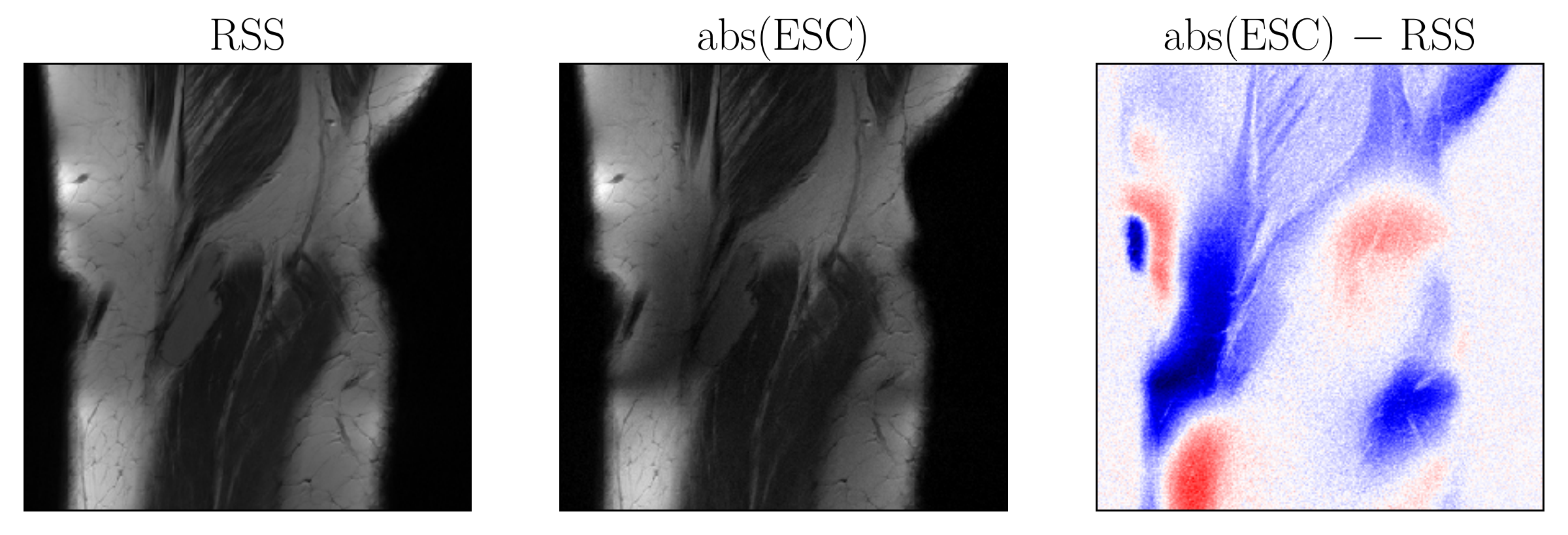}}}
\caption{second ``worst'' abs(ESC)$-$RSS;
``abs'' refers to the absolute value of a complex number}
\label{err2}
\end{figure}

\begin{figure}
\centerline{\parbox{.8\textwidth}{\includegraphics[width=.8\textwidth]{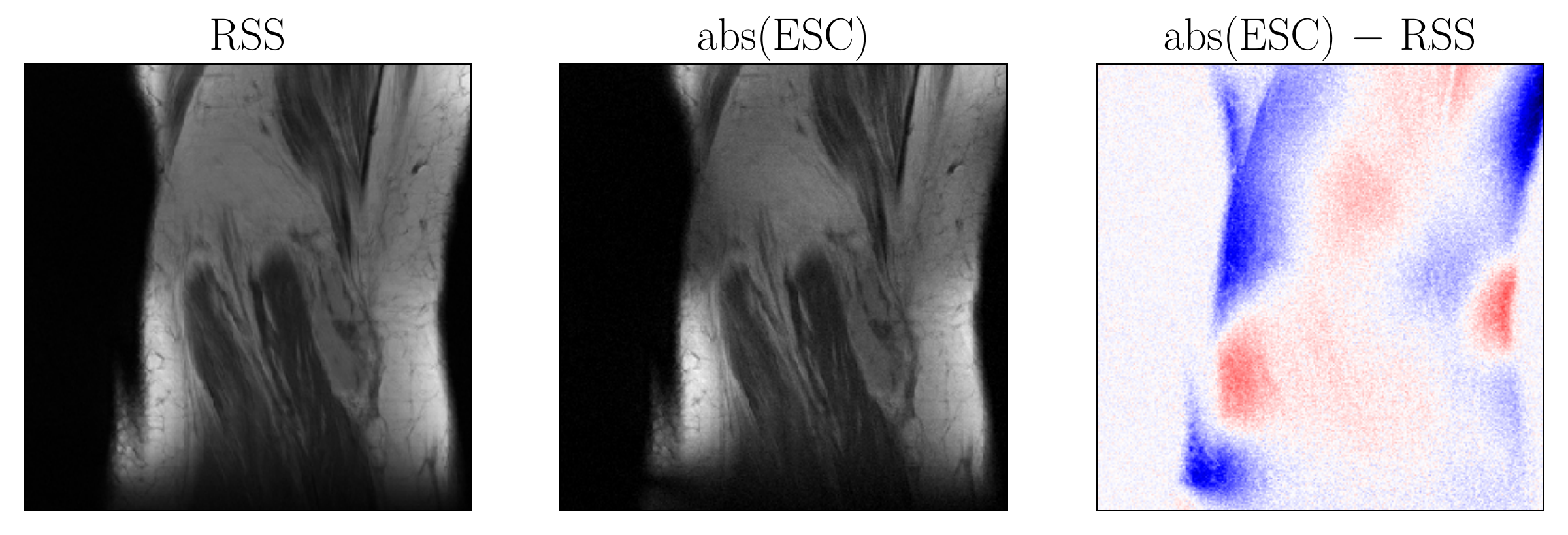}}}
\caption{third ``worst'' abs(ESC)$-$RSS;
``abs'' refers to the absolute value of a complex number}
\label{err3}
\end{figure}

Figures~\ref{err1}--\ref{err3} display examples for which ESC deviates
substantially from the ground-truth RSS. These are among the worst problems
we identified (though of course ``worst'' is somewhat subjective --- SSIM
of~\cite{wang-bovik-sheikh-simoncelli} was helpful in finding bad cases);
about 4\% of the volumes scanned exhibited some indication of such a problem,
with about 10\% of the cross-sectional slices within those volumes
showing signs of the problems displayed in Figures~\ref{err1}--\ref{err3}.
Overall, then, around 0.4\% of the two-dimensional images in the data set
display significant artifacts.
The main culprit seems to be ringing artifacts near the boundary
of the object being imaged.
Whether such occasional ringing oscillations are an artifact
of our fitting procedure or are inherent in the data
when converting from multiple receivers remains unclear;
however, the ``ground-truth'' RSS reconstructions do exhibit similar
(albeit abated) artifacts.

\begin{remark}
The method proposed in the present paper is heavily data-driven.
If computational simulations or excellent empirical estimates
of the ``physical coil sensitivity maps'' defined in~\cite{ohliger-sodickson}
could be calculated with high fidelity to the physical reality
represented in the fastMRI data of~\cite{fastMRI},
then less data-intensive methods might be feasible.
Unfortunately, actual MRI scanners are exceptionally complex
and deviate significantly from textbook idealizations;
interactions of the objects being imaged with the scanners
further distort the sensitivities of receiver coils,
as detailed, for example, by~\cite{lee-pauly-nishimura}.
While autocalibration procedures such as those from ESPIRiT of~\cite{espirit}
work well for multi-coil reconstruction, their estimates of sensitivity maps
are specifically tailored for multi-coil reconstruction,
not for constructing single-coil measurements from multi-coil.
\end{remark}

\clearpage

\clearpage

\section{Conclusion}

Although our simulation of a single-coil scanner might appear perverse
and pointless at first glance --- after all, we throw away
the information necessary to realize the gains in signal-to-noise ratios
that imaging with multiple receivers provides
--- maintaining fidelity to a physically realizable simple MRI machine in fact
yields a convenient test case, a reliable stepping-stone toward
the complexity of full parallel-MRI reconstruction.
Less than 0.5\% of the resulting emulated single-coil (ESC) images exhibit
artifacts such as those displayed in Figures~\ref{err1}--\ref{err3};
Figures~\ref{jzb1}--\ref{jzb6} display typical examples.
The obvious alternative to our scheme would be to take measurements
on an actually manufactured single-coil MRI scanner;
however, MRI is extremely expensive, especially in the medical domain
considered for the fastMRI data set of~\cite{fastMRI}.


\section*{Acknowledgements}

We would like to thank Matt Muckley, Erich Owens, Mike Rabbat, Dan Sodickson,
and Anuroop Sriram.

\bibliography{mri}
\bibliographystyle{siam}

\end{document}